# Immune response to functionalized mesoporous silica nanoparticles for targeted drug delivery

Simon Heidegger[1,2], Stefan Niedermayer[3], Alexandra Schmidt[3], Dorothée Gößl[3], Christian Argyo[3], Stefan Endres[1], Thomas Bein[3,*], Carole Bourquin[1,4,*]

[1] Center for Integrated Protein Science Munich (CIPSM), Division of Clinical Pharmacology, Medizinische Klinik und Poliklinik IV, Ludwig-Maximilians-Universität München, 80336 Munich, Germany

[2] III. Medizinische Klinik, Klinikum Rechts der Isar, Technische Universität München, 81675 Munich, Germany

[3] Department of Chemistry and Center for NanoScience (CeNS), University of Munich (LMU), 81377 Munich, Germany

[4] Chair of Pharmacology, Department of Medicine, Faculty of Science, University of Fribourg, 1700 Fribourg, Switzerland

[*] These authors contributed equally.

Corresponding Authors:

Carole Bourquin, Chair of Pharmacology, Department of Medicine, Faculty of Science, University of Fribourg, 1700 Fribourg, Switzerland (carole.bourquin@unifr.ch);

Thomas Bein, Department of Chemistry and Center for NanoScience (CeNS), University of Munich (LMU), 81377 Munich, Germany (bein@lmu.de)

# ABSTRACT

Multifunctional mesoporous silica nanoparticles (MSN) have attracted substantial attention with regard to their high potential for targeted drug delivery. For future clinical applications it is crucial to address safety concerns and understand the potential immunotoxicity of these nanoparticles. In this study, we assess the biocompatibility and functionality of multifunctional MSN in freshly isolated, primary murine immune cells. We show that the functionalized silica nanoparticles are rapidly and efficiently taken up into the endosomal compartment by specialized antigen-presenting cells such as dendritic cells. The silica nanoparticles showed a favorable toxicity profile and did not affect the viability of primary immune cells from the spleen in relevant concentrations. Cargo-free MSN induced only very low immune responses in primary cells as determined by surface expression of activation markers and release of pro-inflammatory cytokines such as Interleukin-6, -12 and -1β. In contrast, when surface-functionalized MSN with a pH-responsive polymer capping were loaded with an immune-activating drug, the synthetic Toll-like receptor 7 agonist R848, a strong immune response was provoked. We thus demonstrate that MSN represent an efficient drug delivery vehicle to primary immune cells that is both non-toxic and non-inflammagenic, which is a prerequisite for the use of these particles in biomedical applications.

# INTRODUCTION

In recent years the development of nanoparticles for various biomedical applications has been the focus of intense research.[1] Rationally designed nanoparticles are engineered for targeted delivery of various drugs or vaccines.[2] With spatio-temporally controlled release of their therapeutic cargo, such nanoparticles have the potential to increase drug efficacy while minimizing undesired off-target effects. However, due to their nanoscale size, chemical composition and surface reactivity, nanoparticles can be potentially detected by and interact with the host immune response.[3] While in certain applications (e.g. vaccine delivery) an immunostimulatory function may be desirable,[4] uncontrolled systemic immune activation will limit their therapeutic use. Thus, the profound understanding of a nanomaterial's interaction with the immune system and its possible stimulatory and suppressive actions are a critical prerequisite for any clinical application.

Specialized antigen-presenting cells (APC) of the innate immune system, such as dendritic cells (DC), constantly sample their surroundings, taking up cell debris and foreign materials. These cells are equipped with a variety of pattern-recognition receptors that allow for the detection of invading pathogens or signs of cell stress and damage. Ligation of these receptors results in maturation of APCs associated with expression of costimulatory molecules (such as CD80 and CD86), the release of proinflammatory cytokines and eventually the initiation of a subsequent adaptive immune response.[5] DCs have been shown to efficiently engulf various kinds of nanoparticles both in vitro and in vivo,[6,7] but the consequences with regard to subsequent immune responses remain poorly understood.

Previous work of some of us has been focused on the development of a portfolio of core-shell colloidal mesoporous silica nanoparticles.[8,9] Due to their different molecular functionalization in the core and the shell, the MSN can be equipped with internal functionality for controlled host-guest interactions, a system for accurate cargo release upon external stimuli, as well as targeting ligands towards the required type of cell. A broad variety of different triggered-release capping systems were presented in the past years, based on external stimuli such as light[10,11], heat,[12] or magnetic fields,[13] or triggered by intracellular events including a change in pH,[14,15] redox reactions,[16] or the presence of enzymes.[17] For example, we recently reported a multifunctional system based on a pH-responsive polymer (poly(2-vinylpyridine), MSN-PVP) that allows for the facile delivery of membrane-permeable cargos into cancer cells.[18] With the possibility to integrate almost any functionality of interest, as well as the efficient synthesis, this system holds promise for wide-ranging biological and medical applications, especially in cancer therapy.

In this study, we investigate the immune-modulatory properties of MSN in primary murine splenocytes. We show that MSN themselves provoke only a marginal immune response but when loaded with a synthetic immune activator can function as an efficient delivery tool for potent immune activation.

## METHODS

**Materials.** All reagents were purchased from commercial suppliers. Tetraethyl orthosilicate (TEOS, Sigma-Aldrich, >98%), cetyltrimethylammonium chloride (CTAC, Fluka, 25wt% in $H_2O$), triethanolamine (TEA, Sigma-Aldrich, 98%), (3-aminopropyl)-triethoxysilane (APTES, Fluka, 95%), N-(3-dimethylaminopropyl)-N-ethylcarbodiimide hydrochloride (EDC, Sigma-Aldrich, 97%), ammonium nitrate (Sigma-Aldrich, 99%), conc. hydrochloric acid (Sigma-Aldrich, >95%, 37 wt%), α-amino-ω-carboxy terminated poly(2-vinylpyridine) ($NH_2$-PVP-COOH, Polymer Source, $M_n$ = 10.000, PDI = 1.08), Boc anhydride (Aldrich, 95%), N-hydroxysulfosuccinimide sodium salt (*sulfo*-NHS, Sigma-Aldrich, 98%), ethanol (EtOH, Sigma-Aldrich, >99.5%), tetrahydrofuran (THF, anhydrous, Sigma-Aldrich, ≥99.9%), trifluoroacetic acid (TFA, Acros Organics, 99%), magnesium sulfate ($MgSO_4$, anhydrous, Sigma-Aldrich, ≥99.5%), triethylamine (Sigma-Aldrich, ≥99%), fluorescein isothiocyanate (FITC, Fluka, >90%), saline-sodium citrate buffer concentrate (SSC-buffer (20x), Sigma-Aldrich), citric acid buffer solution (citric acid/HCl/NaCl buffer, pH 2, Sigma-Aldrich), TLR7 agonist R848 (Invivogen, Toulouse). All chemicals were used as received without further purification. Doubly distilled water from a Millipore system (Milli-Q Academic A10) was used for all synthesis and purification steps.

**Preparation of shell-functionalized colloidal mesoporous silica nanoparticles.** Polymer-capped mesoporous silica nanoparticles (MSN-PVP) were synthesized as described previously.[8, 18] In brief, colloidal MSN with an amino-functionality (MSN-$NH_2$) were prepared by a delayed co-condensation approach with cetyltrimethylammonium chloride (CTAC) as structure directing agent and tetraethylorthosilicate (TEOS) as primary silica source.[8] In order to introduce the outer shell functionalization, 3-aminotriethoxysilane (APTES) was added during the condensation process. The synthesis was followed by a facile extraction of the organic template from the mesopores. Then, 50 mg of the template-extracted MSN-$NH_2$ were subsequently reacted with a boc-protected bi-functional polymer (COOH-PVP-NHboc) dissolved in THF in an EDC-assisted amidation reaction. The reaction was followed by the deprotection of the polymer with trifluoroacetic acid, yielding the sample MSN-PVP.[18]

**Fluorescein labeling of the MSN.** Fluorescein isothiocyanate (FITC)-labeling of MSN-NH$_2$ was performed following a described procedure.[19] In brief, 50 mg of MSN-NH$_2$ dispersed in ethanol (25 mL) were added to an ethanolic solution of FITC (25 mL, containing 7.4 mg FITC, 0.19 mmol). The suspension was stirred at ambient temperature in the dark for 24 h. The resulting sample MSN-FITC was collected by centrifugation (19.000 rpm, 43.146 RCF, 20 min) and washed three times with ethanol (25 mL each) by subsequent centrifugation and redispersion. After the final centrifugation step the particles were resuspended in absolute ethanol. For cell experiments, 1 mg of the labeled MSN were centrifuged and resuspended in 1 mL phosphate-buffered saline (PBS).

**MSN loading with the synthetic TLR7 agonist R848.** 1 mg MSN-PVP were dispersed in 240 µL of sterile water. To open the pores and enable the uptake of the respective drug molecule, 50 µL of citrate buffer (pH 2) was added, followed by the addition of 10 µL R848 of the stock solution (1 mg/mL), yielding an overall drug concentration of 10 µg/300 µL in the solution. The particles were stirred overnight, centrifuged, and re-dispersed in 700 µL SSC buffer (pH 7) to enable the closure mechanism of the pH-responsible polymer. The resulting particles were washed extensively with SSC buffer (pH 7) and finally redispersed in 1 mL SSC. For incubation with primary cells, MSN were centrifuged and redissolved in complete RPMI medium (see below).

**Characterization of MSN.** Centrifugation was performed using a Sorvall Evolution RC equipped with a SS-34 rotor or an Eppendorf centrifuge 5418 for small volumes. All samples were investigated with an FEI Titan 80-300 transmission electron microscope operating at 300 kV with a high-angle annular dark field detector. A droplet of the diluted MSN solution in absolute ethanol was dried on a carbon-coated copper grid. Nitrogen sorption measurements were performed on a Quantachrome Instruments NOVA 4000e. All samples (15 mg each) were heated to 60 °C for 12 h in vacuum (10 mTorr) to outgas the samples before nitrogen sorption was measured at 77 K. Pore size and pore volume were calculated by a NLDFT equilibrium model of N$_2$ on silica, based on the desorption branch of the isotherms. In order to remove the contribution of the interparticle textural porosity, pore volumes were calculated only up to a pore size of 8 nm. A BET model was applied in the range of 0.05 – 0.20 p/p$_0$ to evaluate the specific surface area of the samples. Dynamic light scattering (DLS) measurements were performed on a Malvern Zetasizer-Nano instrument equipped with a 4 mW He-Ne laser (633 nm) and an avalanche photodiode. The hydrodynamic radius of the particles was determined by dynamic light scattering in ethanolic or aqueous suspension. For this purpose, 100 µL of an ethanolic suspension of MSN particles (ca. 10 mg/mL) was diluted with 3 mL of ethanol or water prior to the measurement. Zeta potential measurements of the samples were performed on a Malvern Zetasizer-Nano instrument equipped with a 4 mW He-Ne laser (633 nm) and an avalanche photodiode. Zeta potential measurements were performed using the add-on Zetasizer titration system (MPT-2) based on diluted NaOH and HCl as titrants. For this purpose, 1 mg of the particles was diluted in 10 mL bi-distilled water. Thermogravimetric analysis was performed on a Netzsch STA 440 C TG/DSC with a heating rate of 10 K / min in a stream of synthetic air of about 25 mL/min. The mass was normalized to 100% at 150 °C for all samples to take into account solvent desorption.

**Mice.** Female C57Bl/6 mice were purchased from Harlan-Winkelmann. Mice were at 6-8 weeks of age at the time of the experiment. Animal studies were approved by the local regulatory agency (Regierung von Oberbayern, Munich, Germany).

**Media, reagents and cell culture.** Single cell suspensions from spleens were obtained by filtering through a 100 µm cell strainer (BD Biosciences, Heidelberg, Germany). Erythrocytes were lysed with ammonium chloride buffer (BD Biosciences). Cells were cultured in complete RPMI (Roswell Park Memorial Institute) 1640 medium (10% fetal calf serum (FCS), 2 mM L-glutamine, 100 µg/mL streptomycin and 1 IU/mL penicillin) at 37 °C in 10% $CO_2$. In some conditions, CpG 1826 (unmethylated CpG sequence-containing oligonucleotides, a TLR9 agonist, 3 µg/mL, from Invivogen, Toulouse, France) was added to the culture. For maximal IL-1β release, DCs were primed with lipopolysaccharide (LPS, a TLR4 agonist, 20 ng/ml, from Invivogen) overnight and ATP (5 mM, from Sigma-Aldrich) was added to the culture 2 hours prior to the analysis. For exposure to nanoparticles, splenocytes were seeded in complete RMPI at a density of $1.0 \times 10^6$ / mL in 96-well tissue culture plates. Silica nanoparticles were added in complete RPMI medium at the indicated concentration. After 18 - 24 hours, cells and culture supernatant were analyzed.

**Quantification of cytokines.** Cell supernatants were analyzed for cytokine secretion by ELISA (R&D Systems or eBioscience) according to the manufacturers' protocol.

**Flow cytometry and apoptosis assay.** Cell suspensions were stained in PBS with 1% FCS. Fluorochrome-coupled antibodies against the surface antigens B220, CD3, CD4, CD8, CD11b, CD11c, CD69, CD80, F4/80 and appropriate isotype controls were purchased from BioLegend. Data were acquired on a FACSCanto II (BD Biosciences) and analyzed using FlowJo software (Tree Star, Ashland, OR). The Annexin V-FITC Apoptosis Detection KIT 1 (BD Biosciences) was used for detection of apoptotic cells. Following surface staining and 2 washing steps, single-cell suspensions were resuspended in the provided buffer, incubated with Annexin V - FITC and propidium iodide (PI) and subsequently analyzed by flow cytometry. Terminal deoxynucleotidyltransferase dUTP nick end labeling (TUNEL staining) was done with the APO-BRDU™ Kit (BD Pharmingen) according to the manufacturer's protocol.

**Confocal Microscopy.** Splenocytes were incubated for 4 hours with fluorescein-tagged NPs, washed and re-suspended in culture medium. 75 nM LysoTracker (Invitrogen) and 3 µg/mL Hoechst dye (Invitrogen) were used for lysosomal and nuclear staining. Stained cells were visualized using a confocal laser scanning microscope (TCS SP5II, Leica).

**Statistics.** All data are presented as mean ± S.E.M. Statistical significance of single experimental findings was assessed with the independent two-tailed Student's t-test. For multiple statistical comparison of a data set the one-way ANOVA test with Bonferroni post-test was used. Significance was set at *p*-values $p < 0.05$, $p < 0.01$ and $p < 0.001$ and was then indicated with an asterisk (*, ** and ***). All statistical calculations were performed using Graphpad Prism (GraphPad Software).

# RESULTS

**MSN particle characteristics.**

As reported previously,[18] the template-free MSN-NH$_2$ show a wormlike pore structure with an average pore size of 3.8 nm and feature a large surface area (1097 m$^2$/g) which is typical for MSN **(Table 1 and Fig. 1)**.

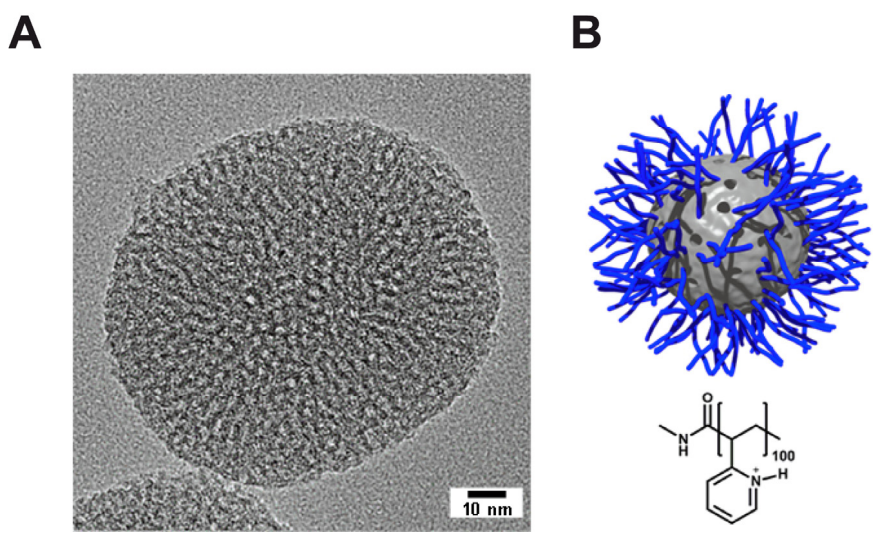

**Figure 1: Structure of mesoporous silica nanoparticles. (A)** Transmission electron micrograph of a template-extracted MSN-NH$_2$, exhibiting a worm-like pore structure. **(B)** Schematic illustration of the pH-responsive nanocarrier system (MSN-PVP) employed in this work at a pH value of 5 (open state). The inorganic-organic hybrid material consists of a mesoporous silica core (grey) and a covalently attached pH-responsive polymer (poly(2-vinylpyridine), blue).

The covalent modification of MSN-NH$_2$ with a boc-protected poly(2-vinylpyridine) and the following de-protection was monitored with several characterization methods; the accessible pore size as determined by nitrogen sorption measurements is barely affected by the surface modification with the pH-responsive polymer, whereas the decrease of the specific surface area is relatively large. This can be explained by the addition of non-porous polymer to the outer surface and the blocking of some pores by frozen polymer on the surface layer of the MSN. Dynamic light scattering (DLS) measurements in aqueous media revealed an average particle size of 160 nm for MSN-NH$_2$ and 550 nm for MSN-PVP. The polymer-modified sample shows some aggregation behaviour in water due to the hydrophobicity at pH 7, indicated by the apparent size increase to 550 nm. However, transmission electron microscopy (TEM) revealed that the polymer-functionalized sample MSN-PVP still features a narrow particle size distribution, which makes them excellent candidates as drug delivery vehicles.

Table 1. Key features of pH-responsive MSN-PVP.

| Sample | Particle size[a] [nm] | BET surface area [m²/g] | Pore size[b] [nm] | Relative mass loss[c] [%] |
|---|---|---|---|---|
| MSN-NH$_2$ | 160 | 1097 | 3.8 | 15 |
| MSN-PVP-NH$_2$ | 550 | 617 | 3.4 | 62 |

[a] Particle size refers to the peak value derived from dynamic light scattering (DLS). [b] Non-linear density functional theory (NLDFT) pore size refers to the peak value of the pore size distribution. [c] Relative mass loss obtained by thermogravimetric analysis (TGA). All curves were normalized to 150 °C.[18]

**Efficient uptake of mesoporous silica nanoparticles by specialized antigen-presenting cells**

To test whether MSN-NH$_2$ can serve as a delivery tool in primary immune cells, freshly isolated mouse splenocytes that harbor a variety of different immune cells were cultured in the presence of fluorescently-labeled MSN (MSN-FITC) overnight. The uptake of labeled MSN-FITC by different cell types was analyzed by flow cytometry. Cells of the innate immune system, that includes macrophages and dendritic cells, showed high uptake of MSN-FITC, as measured by fluorescence signal-positive cells **(Fig. 2A)**. As such, dendritic cells, which are highly specialized in antigen uptake, processing and presentation, were more efficient in uptake than macrophages. In contrast, T and B cells, which are the effector cells of the adaptive immune system, showed only trace fluorescence signal positivity. The intracellular uptake of MSN-FITC was clearly concentration-dependent **(Fig. 2B)**. Fluorescent microscopy showed a speckled distribution pattern of fluorescent signals within dendritic cells, suggesting uptake of labeled MSN into distinct intracellular compartments but not into the cytosol or nucleus **(Fig. 2C)**. Indeed, counter-staining with a fluorescent marker for lyso-/endosomes showed co-localization with the fluorescein-labeled MSN. In summary, MSN show rapid and efficient uptake into the endosomal compartment of specialized antigen-presenting cells such as dendritic cells but not into adaptive immune cells.

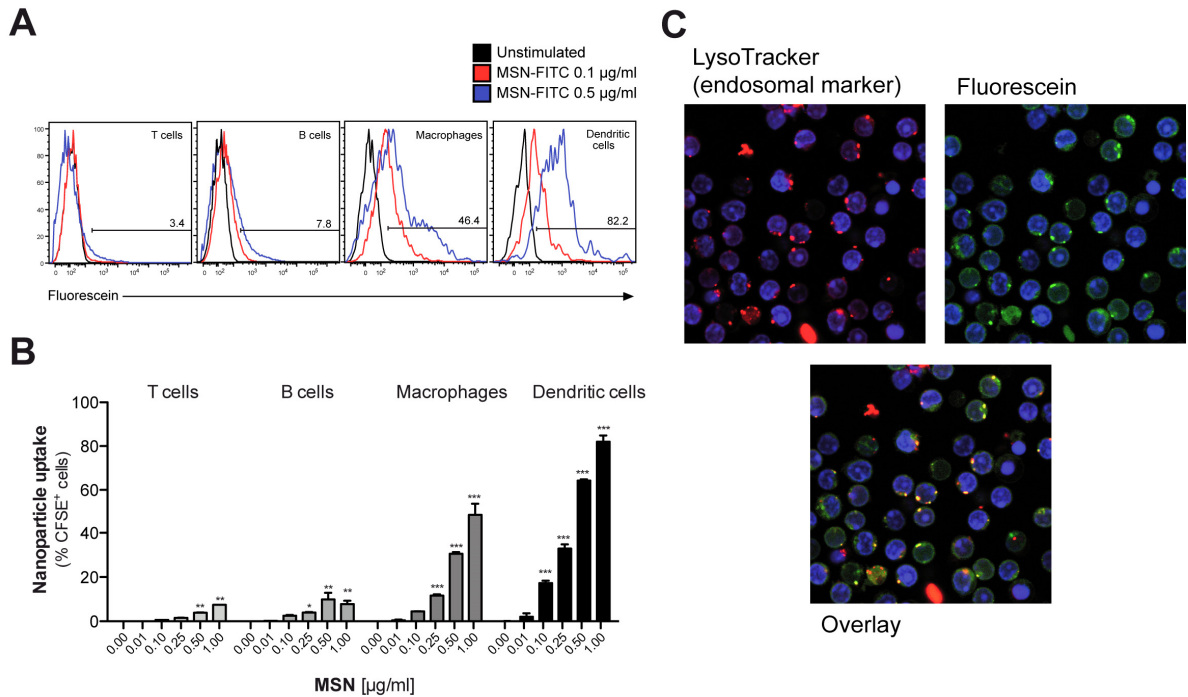

**Figure 2: Mesoporous silica nanoparticles are efficiently taken up by specialized antigen-presenting cells. (A, B)** Freshly isolated, total splenocytes were incubated for 18 h with different concentrations of fluorescein-labeled MSN (MSN-FITC). The uptake of fluorescence signals by different cell populations was determined by flow cytometry. **(A)** Representative histograms are gated on the indicated cell subset. The numbers give the percentage of fluorescein highly positive-stained cells. **(B)** Diagrams show the mean percentage of fluorescein-positive cells of triplicate samples ± s.e.m. An asterisk indicates comparison with unstimulated cells. **(C)** Complete splenocytes were incubated for 3 h with 0.1 μg/mL MSN-FITC. Cell endosomes were stained with LysoTracker™ and intracellular MSN localization was determined by fluorescence microscopy. All results are representative of at least two independent experiments. No stim., no stimulation.

**MSN-NH$_2$ are non-toxic and do not affect the viability of splenocytes, unless used in very high concentrations**

We next sought to determine the cytotoxicity of MSN for primary cells. For this purpose, freshly isolated splenocytes were cultured in the presence of increasing concentrations of MSN-NH$_2$ and 18 hours later, we performed Annexin V / propidium iodide (PI) analysis by flow cytometry. Annexin V binds to phosphatidylserins, which in apoptotic cells are translocated from the inner to the outer leaflet of the plasma membrane, and are thus exposed to the external cellular environment.[20] PI is a small molecule that intercalates into double-stranded DNA and becomes fluorescent upon intercalation. PI can only reach nuclear DNA when the cell's integrity is severely compromised during late apoptosis and cell death. The gating strategy for early and late apoptotic cells is depicted in the representative dot blot **(Fig. 3A)**. MSN-NH$_2$ showed a favorable (low) toxicity profile in primary cells and induced marked apoptosis only when used in very high concentrations of 200 µg mL$^{-1}$. The common cytotoxic chemotherapeutic drug oxaliplatin was used as a positive control. To confirm these data, we also performed terminal deoxynucleotidyltransferase dUTP nick end labeling (TUNEL). During the late phase of apoptosis endonucleases degrade the higher order chromatin structure into small DNA pieces. With the TUNEL assay these DNA fragments can be identified through addition of bromolated deoxyuridine triphosphates (Br-dUTP) to the 3'-hydroxyl (OH) termini of double- and single-stranded DNA by the endogenous enzyme terminal deoxynucleotidyl transferase (TdT) and subsequent staining with an FITC-labeled anti-BrdU antibody. The TUNEL analysis confirmed that MSN-NH$_2$ are non-toxic to primary murine splenocytes over a wide concentration range **(Fig. 3B)**.

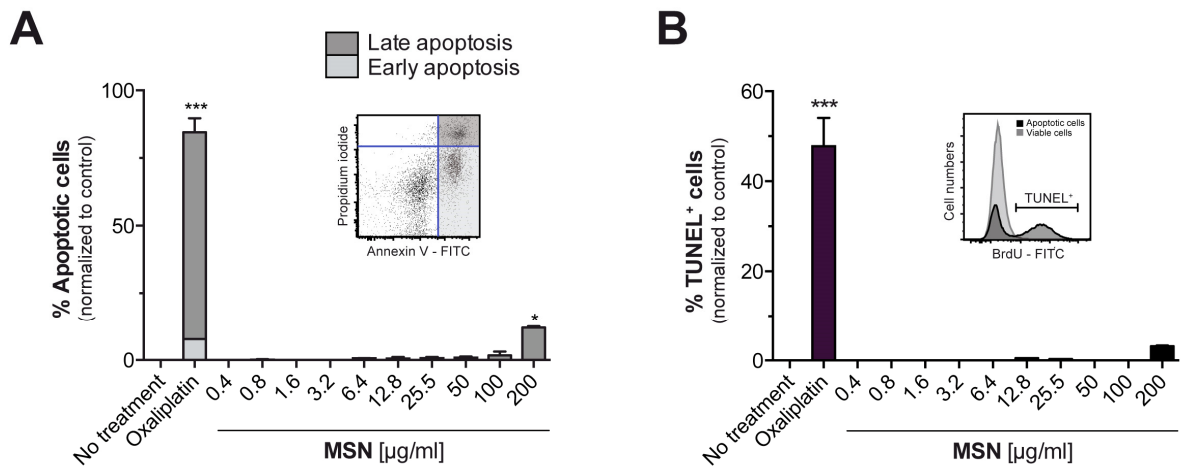

**Figure 3: MSN-NH$_2$ are non-toxic and do not affect the viability of splenocytes, unless used in very high concentrations.** Complete splenocytes were incubated for 18 h with different concentrations of MSN-NH$_2$. **(A)** Cell integrity and viability of splenocytes as determined by propidium iodide (PI) exclusion and Annexin V staining was analyzed by flow cytometry. The dot blot shows the gating strategy of viable (PI$^-$ Annexin V$^-$), early (PI$^-$, Annexin V$^+$) and late apoptotic (PI$^+$, Annexin V$^+$) cells. **(B)** DNA fragmentation was assessed by TUNEL assay. The histogram shows the gating strategy for TUNEL$^+$ cells. All data give the mean percentage of apoptotic cells of triplicate samples ± s.e.m. The mean base-line level of apoptosis in the untreated control group was set as zero %. An asterisk indicates comparison to untreated cells. Results are representative of at least two independent experiments.

**Mesoporous silica nanoparticles induce only very low immunological responses in primary myeloid immune cells**

As a next step, we focused on the immunological response of mammalian primary immune cells to cargo-free MSN. For the implementation of any nanoparticle constructs as molecular delivery system, it is essential to fully understand their immune-modulatory potential in order to tightly control the initiation of immune responses according to the therapeutic goal. Thus, freshly isolated splenocytes were cultured in the presence of MSN-NH$_2$. After 24 hours, surface expression of the B7 family member CD80, a co-stimulatory molecule and activation marker, was examined on different antigen-presenting cells by flow cytometry with specific fluorochrome-coupled antibodies. Additionally, the concentration of the secreted pro-inflammatory cytokines IL-6 and IL-12p70 in the culture supernatant of stimulated cells was quantified by enzyme-linked immunosorbent assay (ELISA). Synthetic unmethylated CpG sequence-containing oligonucleotides (CpG-ODN) that resemble bacterial DNA were used as a positive control. CpG-ODN bind to the endosomal Toll-like receptor (TLR) 9, thereby initiating a full-blown immune response.[21] In comparison to CpG-ODN, MSN-NH$_2$ induced only low levels of CD80 expression on monocytes and dendritic cells, indicating that the cargo-free nanoparticles barely activate immune responses **(Fig. 4A)**. The secretion profile of pro-inflammatory cytokines confirmed these findings, as cells cultured with MSN-NH$_2$ released only low amounts of IL-6 and IL-12p70 **(Fig. 4B)**. Crystalline silica (found in nature as sand or quartz) have been shown to activate a cytosolic multi-protein complex called the NALP3 inflammasome resulting in the release of bioactive IL-1β, a very potent pro-inflammatory cytokine.[22] In contrast, MSN-NH$_2$ only induced trace levels of IL-1β **(Fig. 4C)**. In summary, these data demonstrate that MSN-NH$_2$ only mildly activate primary murine APCs.

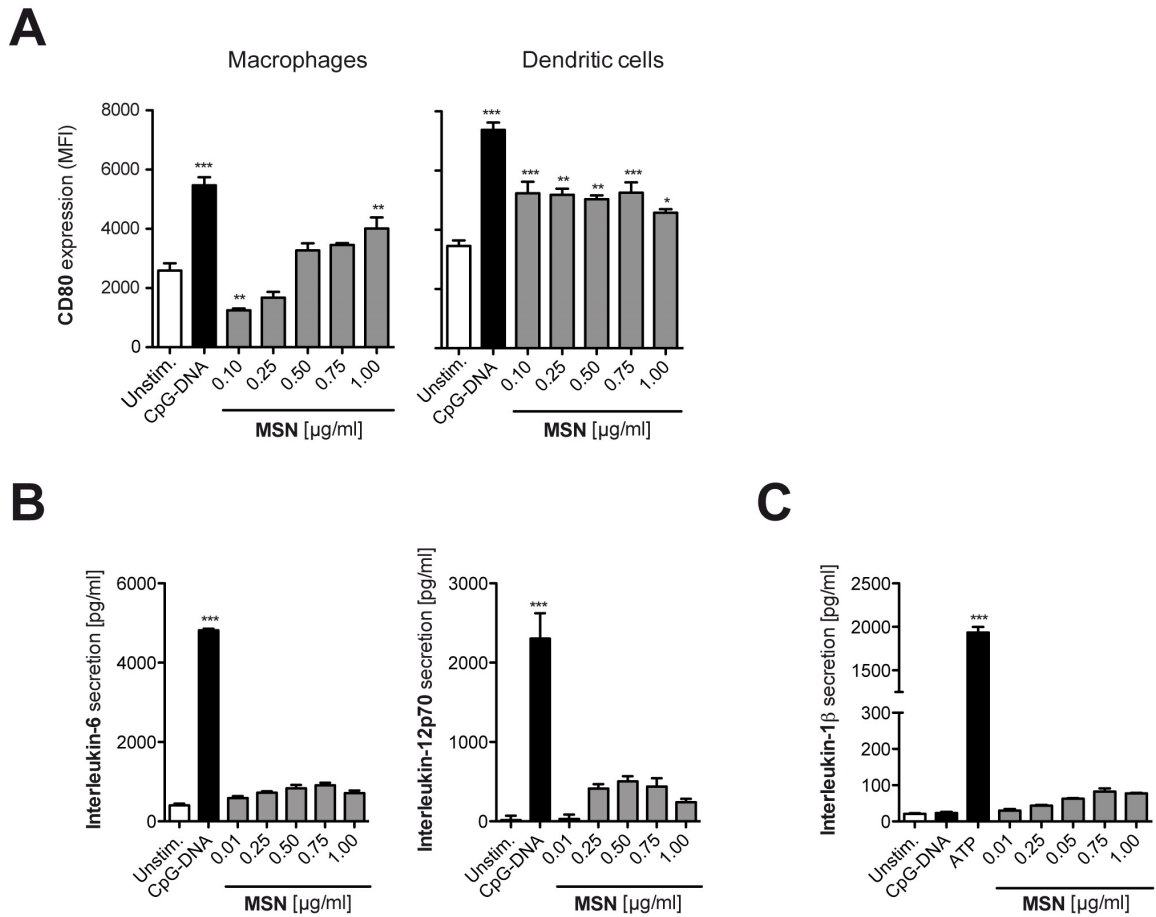

**Figure 4: Mesoporous silica nanoparticles induce only very low immunological responses in primary myeloid immune cells.** Complete splenocytes were incubated for 18 h with different concentrations of MSN-NH$_2$. **(A)** The surface expression of the co-stimulatory molecule CD80 on myeloid antigen-presenting cells was analyzed by flow cytometry. Data give the mean fluorescence intensity (MFI) of the indicated marker on triplicate samples ± s.e.m. The amounts of the pro-inflammatory cytokines **(B)** IL-6, IL-12p70 and **(C)** IL-1β in the cell culture supernatant were determined by ELISA. Data give the mean values of triplicate samples ± s.e.m. An asterisk indicates comparison to unstimulated cells. All results are representative of at least two independent experiments.

**Bystander lymphoid cells are not stimulated by mesoporous silica nanoparticles**

T and B lymphocytes are the effector cells of the adaptive immune system. We have shown that these cell types do not efficiently take up MSN-FITC and are thus unlikely to directly recognize these nanoparticles **(Fig. 3A,B)**. However, T and B cells may react to low levels of pro-inflammatory cytokines released by antigen-presenting cells in response to MSN. In order to investigate the immunostimulatory effect of MSN on such bystander lymphocytes, complete splenocytes (containing both antigen-presenting cells and lymphocytes) were cultured in the presence of MSN-NH$_2$ and expression of the transmembrane C-type lectin

CD69, an early activation marker on B and T cells, was analyzed by flow cytometry. We found that neither B nor T cells showed upregulation of the activation marker CD69 **(Fig. 5)**, indicating that the low-level cytokine release by antigen-presenting cells in response to MSN-NH$_2$ is not sufficient for activation of bystander lymphocytes. In summary, our findings show that MSN-NH$_2$ are rapidly taken up into specialized antigen-presenting cells but are non-toxic and only weakly immunostimulatory to primary murine immune cells.

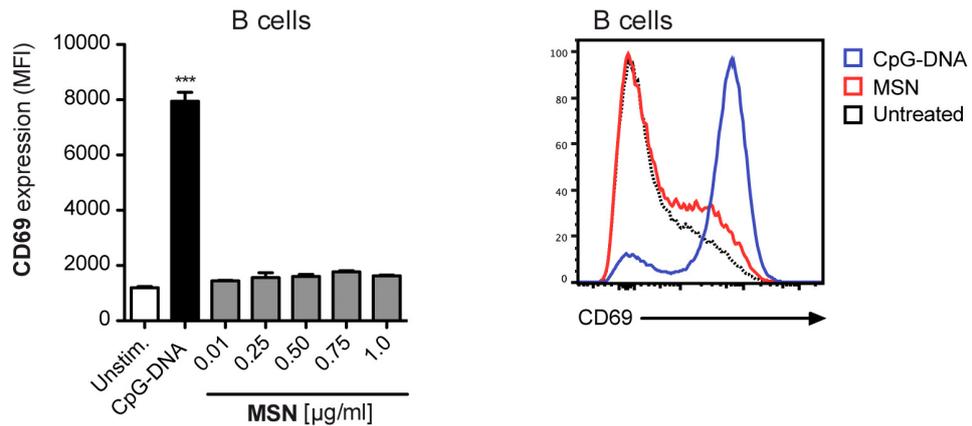

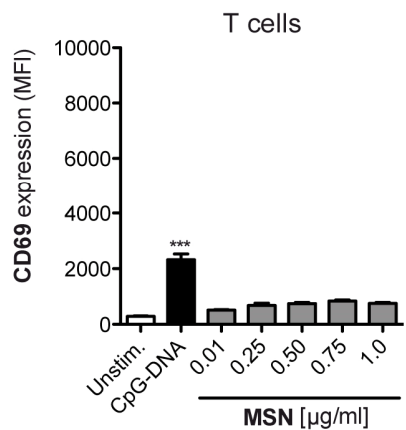

**Figure 5: Mesoporous silica nanoparticles do not result in activation of bystander lymphoid cells.** Complete splenocytes were incubated for 18 h with different concentrations of MSN-NH$_2$. Expression of the activation marker CD69 on effector **(A)** B lymphocytes and **(B)** T lymphocytes was analyzed by flow cytometry. The representative histogram is gated on B220$^+$ B cells and illustrates CD69 expression (black dotted line, unstimulated cells; red line, 1 μg/mL MSN; blue line, CpG-DNA). Data in the graphs give the mean values of triplicate samples ± s.e.m. An asterisk indicates comparison to unstimulated cells. All results are representative of at least two independent experiments.

**Functionalized mesoporous silica nanoparticles are an efficient delivery tool for the synthetic immunostimulatory TLR7 ligand R848**

To test whether MSN can principally function as delivery tool in primary immune cells, MSN coated with a pH-responsive polymer (MSN-PVP) were loaded with R848. A defined amount of the drug was adsorbed at low pH values into the mesopores of MSN-PVP, followed by the subsequent closure of the polymer coat on the mesoporous nanoparticles at pH 7. Following internalization and shuttling into the endosome, the local acidic environment allows for re-opening of the mesopores and release of the cargo. The low molecular weight, synthetic imidazoquinoline compound R848 (also called resiquimod) induces potent immune responses upon uptake and ligation to endosomal TLR 7.[23] Indeed, we found that both unbound (molecules suspended in liquid) as well as MSN-PVP-encapsulated R848 induced activation of dendritic cells with potent upregulation of CD80 and release of pro-inflammatory IL-6 **(Fig. 6)**. These data show that MSN-PVP can be used as a drug delivery tool in primary immune cells. A therapeutic approach to use the targeted release of MSN cargo in the endosome of immune cells in order to target endosomal receptors such as TLR7 with stimuli-responsive capping mechanisms in cancer immunotherapy will be the subject of future studies.

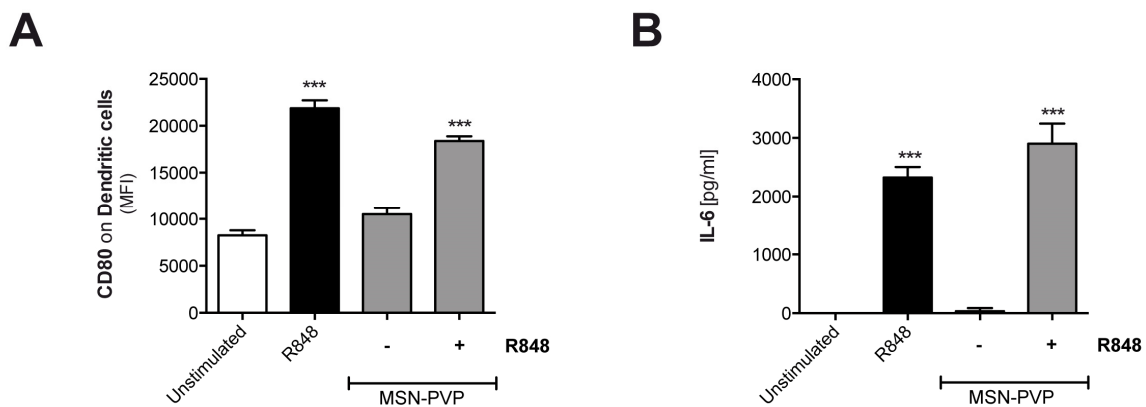

**Figure 6: Mesoporous silica nanoparticles are an efficient delivery tool for the synthetic immunostimulatory TLR7 ligand R848.** Complete splenocytes were incubated for 18 h either with the small molecule immunostimulant R848, cargo-free, or R848-loaded MSN-PVP, respectively. **(A)** The surface expression of the co-stimulatory molecule CD80 on dendritic cells was analyzed by flow cytometry. **(B)** The level of Interleukin-6 in the cell culture supernatant was determined by ELISA. Data give the mean values of triplicate samples ± s.e.m. An asterisk indicates comparison to unstimulated cells. All results are representative of two independent experiments.

## DISCUSSION

Despite intense research in the field of nanomedicine, fundamental knowledge about the interaction of nanomaterials with the cellular components of the host immune system remains scarce. We demonstrate in this work that MSN-NH$_2$ particles are rapidly and efficiently taken up by specialized antigen-presenting cells (APCs) such as dendritic cells and macrophages and are delivered into endo-/lysosomes. Efficient uptake of MSN via endocytosis into non-immune cells, in particular different types of tumor cell lines, has been observed previously.[24, 25, 26] For example, for HeLa cervical adenocarcinoma cells, Mou and co-workers reported a size-dependent endosomal uptake of MSN, favouring a size range of 50 - 120 nm.[27] In contrast to tumor cells, APCs are specialized to scavenge their environment by taking up and sampling cell debris and foreign material, and are also equipped with a variety of germ-line encoded immune receptors to identify invading microorganisms. Many of these receptors, especially several members of the family of Toll-like receptors (TLR), are localized in the endosome and upon ligation of pathogenic material lead to innate immune stimulation. However, our results demonstrate that the uptake of cargo-free MSN-NH$_2$ into APCs did not result in immune activation, as dendritic cells and macrophages showed only mild upregulation of the co-stimulatory molecule CD80 and released only low levels of pro-inflammatory cytokines. Similarly, Lee *et al.* showed that incubation of a macrophage cell line or peritoneal macrophages with MSN resulted in only trace release of cytokines and that short-term *in vivo* application of MSN in mice did not lead to contact hypersensitivity.[28] However, treatment of mice with MSN over a time course of several weeks resulted in histological changes in liver and spleen.[29] We note that such findings are expected to be very dependent on the size, surface functionalization and zeta potential of the particles and thus cannot be generalized for different types of MSN. Generally, the impact of repetitive treatments with MSN must be carefully evaluated before long-term clinical applications are conducted.

Crystalline silica have been shown to potently activate the NLRP3 inflammasome, a cytosolic multiprotein complex, triggering the release of the bioactive form of the potent pro-inflammatory cytokine IL-1β.[30] Similar to crystalline silica, non-functionalized, amorphous silica nanoparticles can activate the inflammasome, leading to significant IL-1β secretion.[31] In contrast, here we demonstrate that the molecularly functionalized MSN-NH$_2$ particles do not induce relevant levels of IL-1β release from primary murine splenocytes. Initial reports suggested that phagocytosis of crystalline silica with a median particle size of 5 µm results in presumably osmotic swelling and damage of the lysosome, leading to activation of the NALP3 inflammasome, which is triggered by lysosomal rupture and content release, not the crystal structure itself.[30] While others linked the formation of reactive oxygen species during an oxidative stress situation to NLRP3 activation,[32] the exact molecular mechanism and prerequisite for inflammasome activation remains to be determined. We propose that due to their size and molecularly functionalized surface structure, spherical MSN-NH$_2$ in comparison to silica crystals do not induce lysosomal damage and subsequent inflammasome-mediated IL-1β release. Generally, besides the physicochemical properties of particle size and surface chemistry, the biological activity of MSN has also been attributed to shape features such as aspect ratio and morphology.[33]

MSN have attracted much interest for their potential as drug delivery vehicles to control various cell functions by the stimuli-responsive delivery of bioactive cargos.[34, 35, 36] An ideal drug delivery vehicle based on MSN may be composed of a multifunctional silica core able to specifically control the interaction with diverse active cargo components. The cargo molecules (e.g. pharmaceutically active drugs) are adsorbed in the mesopores of the nanoparticles, yielding an effective shielding from external degradation in biological fluids.[37, 38] Such multi-functional MSN have been successfully evaluated as antigen carriers and adjuvants for vaccine delivery.[39] Thereby, MSN have shown intrinsic adjuvant activity under certain conditions, thus potentiating antigen-specific T-cell immune responses.[40,41] Interestingly, MSN have been described to enhance MHC class I-restricted presentation of antigens by human dendritic cells.[42] This process called cross-presentation is a vital prerequisite for the induction of adaptive T-cell immunity against exogenous antigens such as tumor proteins. We found that the functionalized MSN-NH$_2$ without protein or adjuvant cargo do not interact with or activate T- and B-lymphocytes *in vitro*. These findings underline the important role of dendritic cells at the interface of innate and adaptive immunity. Importantly, the lack of unspecific lymphocyte priming by the MSN carrier system is a promising requisite for future *in vivo* applications in order to use the high specificity of molecular immunostimulants either on the surface of the MSN or delivered as cargo from its pore system.

Generally, the application of immunostimulatory adjuvants or vaccines via MSN harbors the risk of undesired systemic inflammatory responses upon the uptake of cargo-loaded MSN and subsequent cargo release. These dangerous adverse events can possibly be circumvented by context-dependent, spatiotemporally controlled cargo release. We and other groups pursue a promising approach that takes advantage of internal triggers such as an intracellular change in pH.[14, 15] The efficient pH-responsive closing and opening mechanism of a reversible polymer cap system has been previously demonstrated by time-based fluorescence release experiments; fluorescent dyes were used in these studies.[18, 43] In this work, the synthetic TLR7 agonist R848 (resiquimod) was used as active cargo. R848-loaded MSN-PVP particles induced strong activation of dendritic cells with potent release of pro-inflammatory cytokines. As a defined ligand for endosomal TLR7/8, R848 was presumably released after pH-dependent reopening of the mesopores in the endosome. Whether such spatially controlled release can augment the efficacy and regulation of the subsequent immune response is the focus of ongoing research. Similarly, temporally defined release of the MSN-PVP cargo also appears attractive, as the kinetics of receptor sensitivity strongly influence the outcome of R848-based cancer immunotherapy.[44]

In summary, we demonstrate in this study that MSN-NH$_2$ nanoparticles are non-toxic to primary murine leukocytes and provoke only trace immune activation. In addition, surface functionalized MSN-PVP can serve as a pH-triggered drug delivery tool for the synthetic TLR7/8 ligand R848 to induce potent immune activation in responder cells. The controlled release of their immunomodulatory cargo by otherwise non-immunogenic MSN is a promising tool in future therapies in order to achieve localized immune activation (e.g. in the tumor microenvironment) while preventing undesired, systemic adverse effects.

## ACKNOWLEDGEMENT

This study was supported by the Swiss National Science Foundation (projects 138284 and 310030-156372 to C.B. and National Centre of Competence in Research Bio-Inspired Materials), the Swiss Foundation for Cancer Research (grant KFS-2910-02-2012 to C.B.), the German Research Foundation Graduiertenkolleg 1202 (to C.B., S.E. and S.H.) and Else-Kröner Fresenius Stiftung (to S. H.). S.N., A.S., D.G., C.A. and T.B. thank the German Research Foundation (DFG, SFB 749 and SFB 1032), the Center for NanoScience (CeNS) and the Nano Initiative Munich (NIM) for financial support. S.N. received a Kekulé grant from the Verband der Chemischen Industrie.

## AUTHORSHIP CONTRIBUTION

S.H., A.S., T.B. and C.B. designed the research, analyzed and interpreted the results and prepared the manuscript. S.N. and D.G. designed and synthesized the mesoporous silica nanoparticles. S.H. performed experiments with primary cells. S.E. and C.A. gave methodological support and conceptual advice. T.B. and C.B. guided the study.

The authors declare no financial conflicts of interest.

...